# Comparative Study of Different Guard Time Intervals to Improve the BER Performance of Wimax Systems to Minimize the Effects of ISI and ICI under Adaptive Modulation Techniques over SUI-1 and AWGN Communication Channels

Md. Zahid Hasan (corresponding author)
Dept. of Information & Communication Engineering
University of Rajshahi, Rajshahi, Bangladesh
e-mail: hasan.ice@gmail.com

Md. Ashraful Islam
Dept. of Information & Communication Engineering
University of Rajshahi, Rajshahi, Bangladesh
e-mail: ras5615@gmail.com

Mohammad Reaz Hossain
Assistant Professor, Dept. of Information &
Communication Engineering
University of Rajshahi, Rajshahi, Bangladesh
e-mail: reaz_ice@yahoo.com

Riaz Uddin Mondal
Assistant Professor, Dept. of Information &
Communication Engineering
University of Rajshahi, Rajshahi, Bangladesh
e-mail: riaz_uddin_ru@yahoo.com

*Abstract*— The WIMAX technology based on air interface standard 802-16 wireless MAN is configured in the same way as a traditional cellular network with base stations using point to multipoint architecture to drive a service over a radius up to several kilometers. The range and the Non Line of Sight (NLOS) ability of WIMAX make the system very attractive for users, but there will be slightly higher BER at low SNR. The aim of this paper is the comparative study of different guard time intervals effect for improving BER at different SNR under digital modulation (QPSK, 16-QAM and 64-QAM) techniques and different communication channels AWGN and fading channels Stanford University Interim (SUI-1) of an WIMAX system. The comparison between these effects with Reed-Solomon (RS) encoder with Convolutional encoder ½ rated codes in FEC channel coding will be investigated. The simulation results of estimated Bit Error Rate (BER) displays that the implementation of interleaved RS code (255,239,8) with ½ rated Convolutional code of 0.25 guard time intervals under QPSK modulation technique over AWGN channel is highly effective to combat in the Wimax communication system. To complete this performance analysis in Wimax based systems, a segment of audio signal is used for analysis. The transmitted audio message is found to have retrieved effectively under noisy situation.

I. INTRODUCTION

The experienced growth in the use of digital networks has led to the need for the design of new communication networks with higher capacity. The telecommunication industry is also changing, with a demand for a greater range of services, such as video conferences, or applications with multimedia contents. The increased reliance on computer networking and the Internet has resulted in a wider demand for connectivity to be provided "any where, any time", leading to a rise in the requirements for higher capacity and high reliability broadband wireless telecommunication systems.

Broadband availability brings high performance connectivity to over a billion users' worldwide, thus developing new wireless broadband standards and technologies that will rapidly span wireless coverage. Wireless digital communications are an emerging field that has experienced a spectacular expansion during the last several years. Moreover, the huge uptake rate of mobile phone technology, WiMAX (Wireless Interoperability for Microwave Access) and the exponential growth of Internet have resulted in an increased demand for new methods of obtaining high capacity wireless networks [1].

Worldwide Interoperability for Microwave Access, known as WiMAX, is a wireless networking standard which aims for addressing interoperability across IEEE 802.16 standard-based products. WiMAX defines a WMAN, a kind of a huge hot-spot that provides interoperable broadband wireless connectivity to fixed, portable, and nomadic users. It allows communications which have no direct visibility, coming up as an alternative connection for cable, DSL3, and T1/E1 systems, as well as a possible transport network for Wi-Fi hot-spots, thus becoming a solution to develop broadband industry platforms. Likewise, products based on WiMAX technology can be combined with other technologies to over broadband access in many of the possible scenarios of utilization.

WiMAX will substitute other broadband technologies competing in the same segment and will become an excellent solution for the deployment of the well-known last mile infrastructures in places where it is very difficult to get with other technologies, such as cable or DSL, and where the costs of deployment and maintenance of such technologies would not be profitable. In this way, WiMAX will connect rural areas in developing countries as well as underserved metropolitan areas. It can even be used to deliver backhaul for carrier structures, enterprise campus, and Wi-Fi hot-spots. WiMAX offers a good solution for these challenges because it provides a cost-effective, rapidly deployable solution [2].

Additionally, WiMAX will represent a serious competitor to 3G (Third Generation) cellular systems as high speed mobile data applications will be achieved with the 802.16 specification.

WiMAX is especially popular in wireless applications because of its resistance to forms of interference and degradation (multipath and delay spread). In short, WiMAX delivers a wireless signal much farther with less interference than competing technologies. However, the higher the number of bits per symbol, the more susceptible the scheme is to intersymbol interference (ISI) and noise. Intersymbol interference occurs due to time dispersion when in a multipath environment, the time delay between the various signal paths is a significant fraction of the transmitted signal's symbol period, a transmitted symbol may arrive at the receiver during the next symbol period and cause intersymbol interference (ISI). At higher data rates, the symbol time is shorter; hence, it takes only a smaller delay to cause ISI. This makes ISI a bigger concern for broadband wireless and mitigating it more challenging. Equalization is the conventional method for dealing with ISI but at high data rates requires too much processing power. OFDM has become the solution of choice for mitigating ISI in broadband systems, including WiMAX to add different guard intervals. Moreover, in the digital implementation of WiMAX, the ISI can be completely eliminated through the use of a guard time intervals [3]. And what modulation technique is used in WiMAX systems is mandatory. Generally the signal-to-noise ratio (SNR) requirements of an environment determine the modulation method to be used in the environment. QPSK is more tolerant of interference than either 16-QAM or 64-QAM [4].

To conclude, guard interval gives two fold advantages, first occupying the guard interval, it removes the effect of ISI (Inter Symbol Interference) and by maintaining orthogonality it completely removes the ICI (Inter Carrier Interference) [5]. Thus a good system design will make the guard interval as short as possible while maintaining sufficient multipath protection. For any duplexing, all SSs (Subscriber Stations) shall acquire and adjust their timing such that all uplink OFDM symbols arrive time coincident at the BS to a accuracy of ±25% of the minimum guard-interval or better [6].

## II. SIMULATION MODEL

This structure corresponds to the physical layer of the WiMAX/IEEE 802.16 WirelessMAN-OFDM air interface. In this setup, The input binary data stream obtained from a segment of recorded audio signal is ensured against errors with forward error correction codes (FECs) and interleaved.

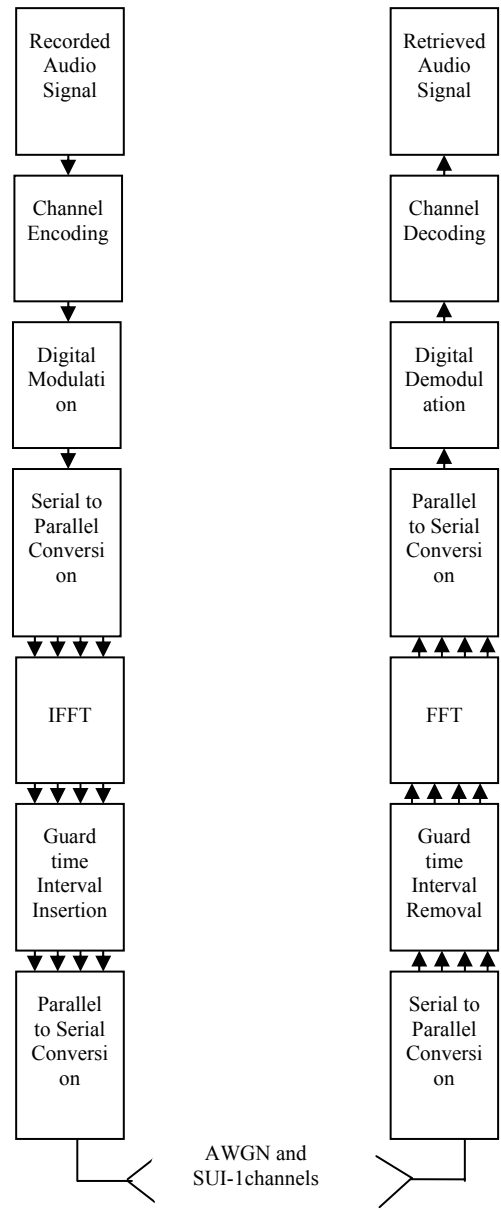

**Figure 1: A block diagram represents Wimax communication system with interleaved concatenated channel coding.**

The complementary operations are applied in the reverse order at channel decoding in the receiver end. The complete channel encoding setup is shown in above Figure 1.

The complementary operations are applied in the reverse order at channel decoding in the receiver end. The complete channel encoding setup is shown in above Figure 1.

FEC techniques typically use error-correcting codes (e.g., RS, CC) that can detect with high probability the error location. These channel codes improve the bit error rate performance by adding redundant bits in the transmitted bit stream that are employed by the receiver to correct errors introduced by the channel. Such an approach reduces the signal transmitting power for a given bit error rate at the expense of additional overhead and reduced data throughput (even when there are no errors) [7]. The forward error control (FEC) consists of a Reed-Solomon (RS) outer code and a rate-compatible Convolutional Code (CC) inner code. A block Reed Solomon (255, 239, 8) code based on the Galois field GF ($2^8$) with a symbol size of 8 bits is chosen that processes a block of 239 symbols and can correct up to 8 symbol errors calculating 16 redundant correction symbols. Reed Solomon Encoder that encapsulates the data with coding blocks and these coding blocks are helpful in dealing with the burst errors [8]. The block formatted (Reed Solomon encoded) data stream is passed through a convolutional interleaver. Here a code rate can be defined for convolutional codes as well. If there are k bits per second input to the convolutional encoder and the output is n bits per second, the code rate is k/n. The redundancy is on not only the incoming k bits, but also several of the preceding k bits. Preceding k bits used in the encoding process is the constraint length m that is similar to the memory in the system [9], where k is the input bits and n is the number of output bits – is equal to ½ and the constraint length m of 7. The convolutionally encoded bits are interleaved further prior to convert into each of the either three complex modulation symbols in QPSK, 16-QAM, and 64-QAM modulation and guard intervals is added to the data once the data is converted into time domain and ready to be transmitted. The addition of guard interval to the data before it is actually transmitted helped the data to cater the problems related to the multipath propagation and provided a resistance against ISI. IEEE 802.16 allows the insertion of guard time intervals of various lengths such 0.25, 0.125, 0.0625 and 0.03125 is added to the WiMAX symbol before it is transmitted. Guard interval is a copy of the last part of OFDM symbol which is appended to the front of transmitted OFDM symbol [10]. The length of the guard interval must be chosen as longer than the maximum delay spread of the target multipath environment. The transmitted data is then fed into the SUI-1 and AWGN channels. At the receiver side, guard interval is removed before any processing starts. As long as the length of guard interval is larger than maximum expected delay spread, all reflections of previous symbols are removed and orthogonality is restored. The orthogonality is lost when the delay spread is larger than length of guard interval.

The simulated coding, modulation schemes and also noisy fading channels used in the present study is shown in Table 1.

**Table 1: Simulated Coding, Modulation Schemes and noisy channels**

| Parameters | Value |
|---|---|
| BW- This is the Nominal Channel Bandwidth | 2.5 MHz |
| $N_{used}$- Number of used subcarriers | 200 |
| G- Ratio of guard time to the useful time | 0.25, 0.125, 0.0625, 0.03125 |
| $N_{FFT}$ | 256 |
| Modulation | QPSK, 16-QAM, 64-QAM |
| RS Code | (255,239,8) |
| CC Code | ½ |
| Noise Channels | AWGN and SUI-1 |
| SUI-1 Channel Model Parameters | |
| P (Power in each tap in dB) | [0  -15  -20] |
| K (Ricean K-factor in linear scale) | [4  0  0] |
| Tau (Tap Delay) | [0  0.4  0.9] |
| Doppler (Doppler maximal frequency parameter) | [0.4  0.3  0.5] |
| Ant_corr (Anteena Correlation) | 0.7 |
| Fnorm (gain normalization factor) | -0.1771 |

In OFDM modulator, the digitally modulated symbols are transmitted in parallel on subcarriers through implementation as an Inverse Discrete Fourier Transform (IDFT) on a block of information symbols followed by an analog-to-digital converter (ADC). To mitigate the effects of inter-symbol interference (ISI) caused by channel time spread, each block of IDFT coefficients is typically presented by a different length of guard intervals. At the receiving side, a reverse process (including deinterleaving and decoding) is executed to obtain the original data bits. As the deinterleaving process only changes the order of received data, the error probability is intact. When passing through the CC-decoder and the RS-decoder, some errors may be corrected, which results in lower error rates [11].

### III.  SIMULATION RESULTS

In this section, we have presented various BER vs. SNR plots for all the essential modulation and coding profiles in the standard on different channel models obtained from simulation experiments. We analyzed audio signal to transmit or receive data as considered for real data measurement. Here uses 8 bits per sample at a sampling frequency of 8 kHz. Now we will present the simulation results both in terms of validation of implementation and values for various parameters that characterize the performance of the physical layer. The main procedure also contains initialization parameters, input data and delivers results. The parameters that can be set at the time of initialization are the number of simulated OFDM symbols, guard interval length, modulation and coding rate, range of SNR values and (SUI-1 and AWGN) channel models for simulation. Figure 2 to 4 shows the BER plots for different guard period and modulation schemes as SNR values are changed on SUI-1 and AWGN channel models. In this simulation, QPSK, 16-QAM, 64-QAM modulation and the varying Guard Intervals length 0.25, 0.125, 0.0625 and 0.03125 is taken. The symbols are transmitted and calculate the Bit Error Rate (BER) performance versus Signal to Noise Ratio (SNR) value from the received signals.

Figure 2, 3 and 4 display the Bit Error Rate (BER) plot obtained in the performance analysis showed that model works well on Signal to Noise Ratio (SNR) less than 14 dB. Simulation results in figure 2 shows the advantage of considering a ½ rated convolutinal coding and Reed-Solomon coding for each of the three considered digital modulation schemes (QPSK, 16-QAM and 64-QAM). The performance of the system of adding 0.25 guard interval under QPSK modulation over AWGN channel is quite satisfactory as compared to other modulation techniques in AWGN and SUI-1 channel.

In figure 2, The Bit Error Rate under QPSK modulation technique over AWGN channel with 0.25 guard time interval for a typical SNR value of 4 dB is 0.000021 which is almost 0 and the SUI-1 channel performance is almost same and smaller than that of 16-QAM and 64-QAM modulations over AWGN and SUI-1 channel.

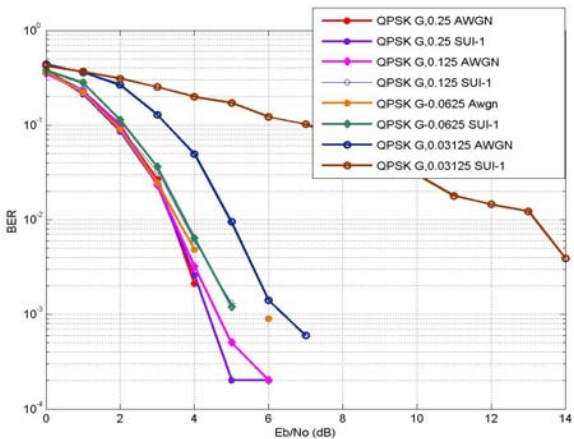

**Figure 2: Comparative study between different guard time intervals of an ½-rated RS-CC coded signal using QPSK modulations under AWGN and SUI-1 channel**

In figure 3 with SUI-1 channel model, the BER performance of 0.25 guard intervals in case of 16-QAM modulation in ½ convolutinal code rate is found no to be suitable for transmission but better performance is shown for QPSK modulation. It is also shown in this figure that the performance of 16-QAM with adding different guard intervals is significant with respect to SNR value.

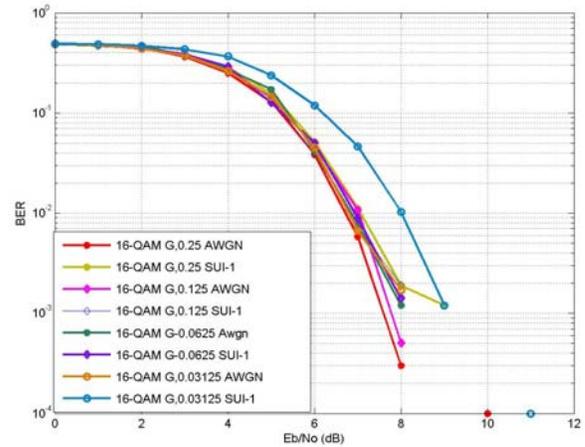

**Figure 3: Comparative study between different guard time intervals of an ½-rated RS-CC coded signal using 16-QAM modulations under AWGN and SUI-1 channel**

In figure 4, The Bit Error Rate under 64-QAM modulation technique over AWGN channel with 0.25 guard time interval for a typical SNR value of 12 dB is almost 0 and the SUI-1 channel performance is same and there is a little difference existed between 64-QAM 0.125 and 64-QAM 0.0625 AWGN communication channel. In table 2 shows that the calculation of approximate Eb/No level in dB with respect to BER at $10^{-3}$ using adaptive modulation techniques with different guard time intervals under AWGN and SUI-1 channel.

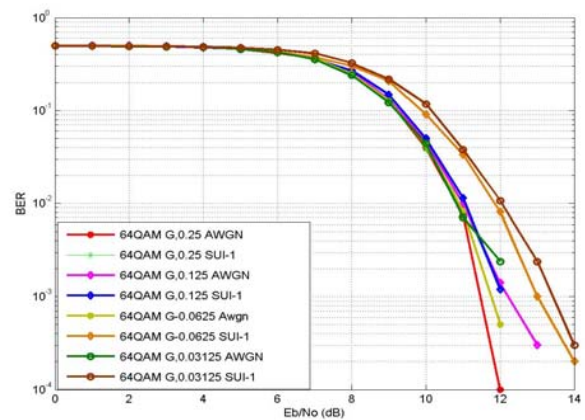

**Figure 4: Comparative study between different guard time intervals of an ½-rated RS-CC coded signal using 64-QAM modulations under AWGN and SUI-1 channel**

**Table 2: Calculate Eb/No in dB with BER at $10^{-3}$ using adaptive modulation techniques with different guard intervals**

| | | | | | | | | |
|---|---|---|---|---|---|---|---|---|
| **QPSK** | | | | | | | | |
| Modulation with different guard intervals | G-0.25 AWGN | G-0.25 SUI-1 | G-0.125 AWGN | G-0.125 SUI-1 | G-0.0625 AWGN | G-0.0625 SUI-1 | G-0.03125 AWGN | G-0.03125 SUI-1 |
| $E_b/N_0$ in dB with BER at $10^{-3}$ | 4 | 4.5 | 4.9 | 5 | 5 | 5.1 | 6 | 14 |
| **16-QAM** | | | | | | | | |
| Modulation with different guard intervals | G-0.25 AWGN | G-0.25 SUI-1 | G-0.125 AWGN | G-0.125 SUI-1 | G-0.0625 AWGN | G-0.0625 SUI-1 | G-0.03125 AWGN | G-0.03125 SUI-1 |
| $E_b/N_0$ in dB with BER at $10^{-3}$ | 7.8 | 8.3 | 7.9 | 8.3 | 8.1 | 8.2 | 8.1 | 9.2 |
| **64-QAM** | | | | | | | | |
| Modulation with different guard intervals | G-0.25 AWGN | G-0.25 SUI-1 | G-0.125 AWGN | G-0.125 SUI-1 | G-0.0625 AWGN | G-0.0625 SUI-1 | G-0.03125 AWGN | G-0.03125 SUI-1 |
| $E_b/N_0$ in dB with BER at $10^{-3}$ | 11.8 | 12 | 12 | 12 | 11.9 | 13 | 12.5 | 13.2 |

The transmitted and received audio signal for such a case corresponding with time and amplitude coordinates is shown in fig5.

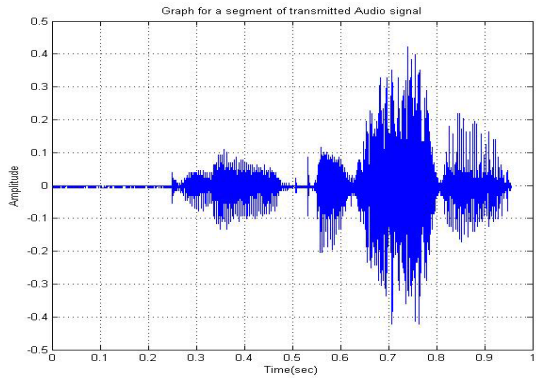

(a)

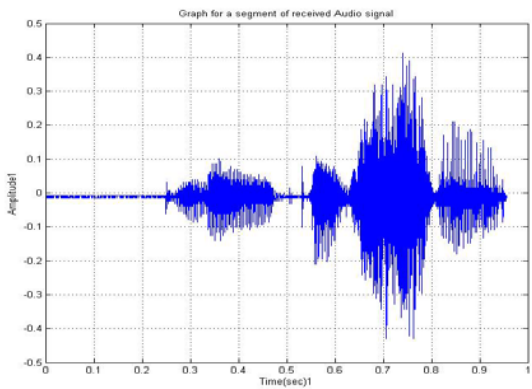

(b)

**Figure 5: A segment of an audio signal, (a) Transmitted (b) Retrieved**

## IV. CONCLUSION AND FUTURE WORKS

A comparative analysis of an Wimax (Worldwide Interoperability for Microwave Access) system adopting concatenated Reed-Solomon and Convolutional encoding with block interval has been carried out. The BER curves were used to compare the performance of different guard time intervals with modulation and coding scheme used. The effects of the different guard intervals over different communication channels (AWGN and SUI-1) were also evaluated in the form of BER. Performance results highlight the impact of modulation scheme and show that the implementation of an 0.25 guard time interval over AWGN interleaved Reed-Solomon with ½ rated convolutional code under QPSK modulation technique provides satisfactory performance among the three considered modulations and there is a little difference existed between other guard time intervals over AWGN and SUI-1 channel.

The IEEE 802.16 standard comes with many optional PHY layer features, which can be implemented to further improve the performance. The optional Block Turbo Coding (BTC) can be implemented to enhance the performance of FEC. Space Time Block Code (STBC) can be employed to provide transmit diversity.